\documentclass[12pt,draftcls,letter,onecolumn]{IEEEtran}
\usepackage[T1]{fontenc} 
\usepackage[utf8]{inputenc} 
\usepackage[english]{babel} 
\usepackage{lipsum} 
\usepackage{url} 
\usepackage{pgfplots}
\pgfplotsset{compat=newest}
\usetikzlibrary{plotmarks}
\usetikzlibrary{arrows.meta}
\usepgfplotslibrary{patchplots}
\usepackage{grffile}

\usepackage{setspace}

\usepackage{amsmath}
\usepackage{amsthm}
\usepackage{bm}
\usepackage{mathtools} 
\DeclarePairedDelimiter\floor{\lfloor}{\rfloor}

\usepackage{authblk}

\usepackage[font=footnotesize]{subfig} 
\usepackage{graphicx}
\graphicspath{{./figures/}}
\usepackage{epstopdf} 
\usepackage[font=footnotesize]{caption}

\usepackage{listings}

\usepackage{booktabs}
\usepackage{tabulary}
\usepackage{caption}
\usepackage{multirow}
\usepackage{multicol}
\usepackage{amssymb}

\usepackage[nolist,nohyperlinks]{acronym}
\begin{acronym}
	\acro{NMA}{navigation message authentication}
	\acro{FEC}{forward error correction}
	\acro{FEA}{forward estimation attack}
	\acro{AN}{artificial noise}
	\acro{SNR}{signal to noise ratio}
	\acro{AWGN}{additive white Gaussian noise}
	\acro{SCER}{security code estimation and replay}
	\acro{iid}{independent and identically distributed}
	\acro{PDF}{probability density function}
	\acro{PSD}{power spectral density}
	\acro{PSK}{phase shift keying}
	\acro{BPSK}{binary phase shift keying}
	\acro{CDF}{cumulative distribution function}
\end{acronym}

\usepackage[inline]{enumitem}

\newcommand{\sigb}{\sigma_{w_B}^2}
\newcommand{\sige}{\sigma_{w_E}^2}
\newcommand{\sigq}{\sigma_{w_q}^2}
\newcommand{\sigs}{\sigma_{w^*}^2}
\newcommand{\sigx}{\sigma_{x}^2}

\newcommand{\sigqe}{\sigma_{w_{q,\epsilon}}^2}
\newcommand{\E}[1]{\mathbb{E} \left[ #1 \right]}

\newcommand{\ie}{i.e., }

\begin{document}
\author{Francesco Formaggio and Stefano Tomasin 
\small \\
Department of Information Engineering \\
University of Padova, Italy \\
\{formaggiof, tomasin\}@dei.unipd.it}

\title{Authentication of Satellite Navigation Signals by Wiretap Coding and Artificial Noise
\footnote{This work has been submitted to the IEEE for possible publication.  Copyright may be transferred without notice, after which this version may no longer be accessible.}}
\date{\today}
\maketitle

\begin{abstract}
In order to combat the spoofing of global navigation satellite system (GNSS) signals  
we propose a novel approach for satellite signal authentication based on information-theoretic security. In particular we superimpose to the navigation signal an authentication signal containing a secret message corrupted by \ac{AN}, still transmitted by the satellite.
We impose the following properties: 
\begin{enumerate*}[label=\emph{\alph*})]
\item the authentication signal is synchronous with the navigation signal,
\item the authentication signal is orthogonal to the navigation signal and
\item the secret message is undecodable by the attacker due to the presence of the \ac{AN}.
\end{enumerate*}
The legitimate receiver synchronizes with the navigation signal and stores the samples of the authentication signal with the same synchronization. After the transmission of the authentication signal, through a separate public asynchronous authenticated channel (e.g., a secure Internet connection) additional information is made public allowing the receiver to 
\begin{enumerate*}[label=\emph{\alph*})]
	\item decode the secret message, thus overcoming the effects of \ac{AN}, and
	\item verify the secret message.
\end{enumerate*}  
We assess the performance of the proposed scheme by the analysis of both the secrecy capacity of the authentication message and the attack success probability, under various attack scenarios. A comparison with existing approaches shows the effectiveness of the proposed scheme.
\end{abstract}
\acresetall

\newpage

\section{Introduction}

\label{sec:intro}
Global navigation satellite system (GNSS) offers positioning and timing services to an increasing variety of users and applications.
The spoofing of GNSS designates the induction of false ranging measurements to a legitimate user by an attacker. 
One of the simplest spoofing techniques is \emph{meaconing}, i.e., the interception and re-broadcast of navigation signals so that the victim computes the ranging estimate based on the spoofer location. More sophisticated versions of this attack selectively forge delayed versions of the ranging signals so that the spoofer can induce any position estimate at the victim.

Detection techniques proposed in \cite{difference1} \cite{difference2} exploit differences between the satellite and the spoofed signal, while \cite{AGC} exploits the receiver's automatic gain control to detect sudden variations of received power due to a spoofing attack.
Using multiple antennas at the receiver can also increase the detection performances \cite{MultiFirst} by estimating the angle of arrival of the received signals \cite{MultiCots,MultiCots2}: this forces the spoofer to not only compute selective delays for each transmitted signal, but also know the angles of arrival and departure of the signal.
A further option to make spoofing attacks more difficult is the inclusion into GNSS signals of data that are (partially) unpredictable at the receiver: in this case the attacker needs  to predict the data signals in order to produce a counterfeit signal. These defence strategies come under the name of \ac{NMA} which aims at authenticating the navigation data, so that the receiver can verify that the received signals come from the legitimate satellite. This mechanism is based on cryptography, and proposals include both symmetric-key  \cite{crypt1,crypt2} and asymmetric-key  \cite{asym1,asym2} encryption. However, typically the data signal includes \ac{FEC} redundancy bits that ease the prediction of  the data codeword from its partial observation. This leads to the \ac{FEA} introduced in \cite{Fea2} and further analysed in \cite{FeaGianluca}. Moreover, encrypted chip values can be estimated and replayed based on received samples \cite{SCER}.

Authentication can be performed also at the physical layer. A general analysis and evaluation framework is presented in \cite{phyLaAuth}, while a recent overview of physical layer authentication methods can be found 
in \cite{tomasinSurvey}.
Typically, authentication methods are classified into 
key-based and key-less methods: in key-less methods users are 
authenticated by verifying that messages of the same user are 
transmitted through the same (initially authenticated) channel \cite{baracca,ferrante}.
An \ac{AN} aided message authentication code is proposed where the \ac{AN} is quantized and transmitted above the physical layer.

In this paper we propose a novel approach for the authentication of satellite navigation signals based on information-theoretic security. We propose to superimpose to the navigation signal an authentication signal corrupted by \ac{AN}, still transmitted by the satellite with the following properties:
\begin{enumerate*}[label=\emph{\alph*})]
\item the authentication signal is synchronous with the navigation signal,
\item the authentication signal is orthogonal to the navigation signal and
\item the secret message is undecodable by the attacker due to the presence of the \ac{AN}.
\end{enumerate*}
In \cite{anaided} \ac{AN} is also used on top of an authentication tag, but no requirements are asked on synchronism and  the system model is different. In our model, instead,
the legitimate receiver synchronizes with the navigation signal and stores the samples of the authentication signal with the same synchronization. After the transmission of the authentication signal, through a separate public asynchronous authenticated channel (e.g. a secure Internet connection) both the \ac{AN} and the secret message are provided  to users who can 
\begin{enumerate*}[label=\emph{\alph*})]
	\item decode the authentication signal, thus overcoming the effects of \ac{AN}, and
	\item verify the content of the authentication message thus authenticating it.
\end{enumerate*} 
Our scheme is based on a key as users are authenticated by verifying a shared secret (the authentication message). However, we perform the key-sharing process after the message has been received. With respect to \cite{anaided} we  apply \ac{AN} at the physical layer and we cancel it before decoding leveraging information-theoretic security approaches.
With respect to \cite{eusipco} that still proposes to superimpose an AN-corrupted authentication signal we include here coding within the framework of wiretap coding.

Moreover, key-based authentication schemes prevent reply attacks by using timestamps.
In the navigation context no reliable timing is available before authentication, therefore time-stamping is not viable.
Moreover, the navigation message is basically already known at the receiver and we must indeed authenticate its timing.
Therefore, we synchronize the navigation and authentication message components so that a spoofing signal will be asynchronous w.r.t. the authentication component, thus directly authenticating the timing.

We analyse the performance of the proposed scheme using information-theoretic security tools for a navigation signal received from a single satellite. We obtain the number of unpredictable bits per transmitted symbol of the secret message even when the spoofer has access to a noiseless navigation signal (still corrupted however by \ac{AN}). The impact of synchronization errors (due to an ongoing spoofing attack) on the authentication system is discussed, and codeword prediction attacks are analysed in terms of success probability. Numerical results are presented showing the effectiveness of the proposed  authentication technique against spoofing attacks also considering various chip pulses for both navigation and authentication spread-spectrum signals.

The rest of the paper is organized as follows. Section \ref{sec:SM} introduces both the system model and the attack strategies. In Section \ref{sec:ITA} we propose the novel authentication protocol, whose design and performance analysis are considered in Section \ref{sec:performanceAnalysis}. Numerical results to support the authentication solution are presented is Section \ref{sec:results} before conclusions are driven in Section \ref{sec:conclusions}.

\section{System Model}
\label{sec:SM}
\begin{figure}
\centering
\includegraphics[width=1.0\columnwidth]{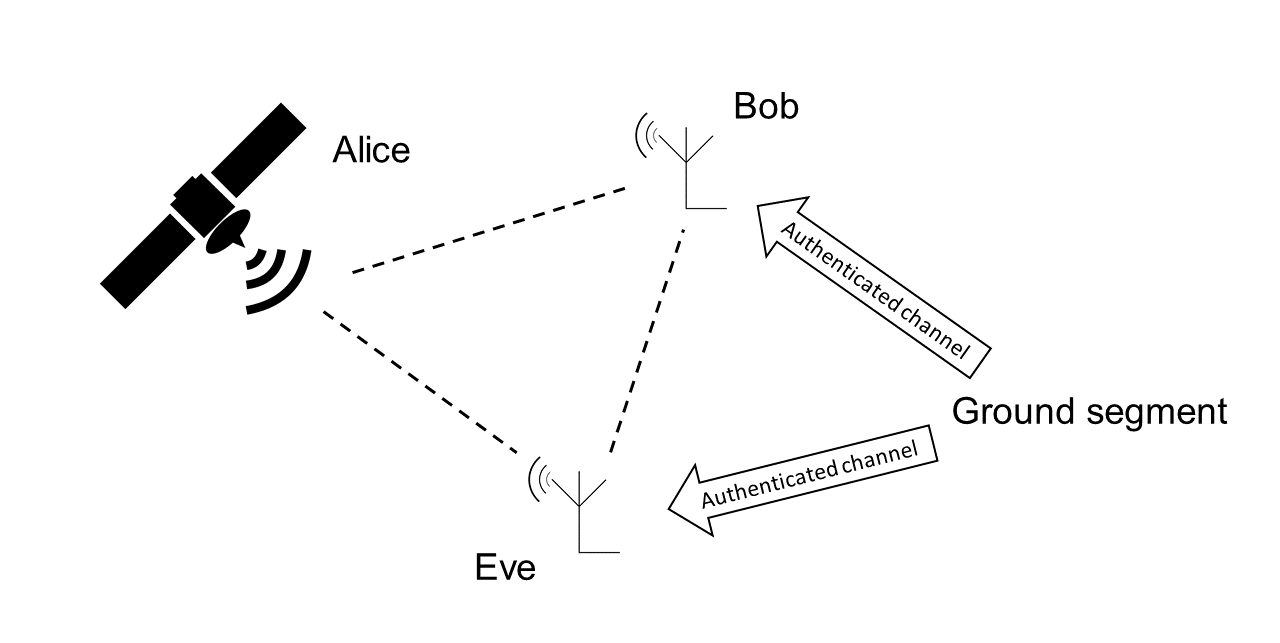}\caption{Reference scenario.}
\label{fig:satelite}
\end{figure}

Fig.  \ref{fig:satelite} shows our reference scenario with a single satellite. Existing systems such as GPS and GALILEO include a satellite (Alice) offering positioning services via a broadcast transmission
to both the legitimate receiver (Bob) and the spoofer (Eve), both assumed to be on the earth. In a spoofing attack Eve transmits a signal that mimics that of the satellite Alice to induce Bob to estimate a wrong position or timing. We also have a forth entity, represented by the {\em ground segment}, i.e., the navigation control center on the earth that is controlling the satellite and can legitimately modify the navigation or authentication signal.

In particular we focus on Galielo GNSS \cite{galileoICT} for civil use transmitted in the E1 band.
The E1 band comprises two signals, data and pilot, added together and distinguishable thanks to different pseudo-random spread-spectrum sequences called \emph{ranging codes} \cite{galileoICT}.
We focus here on the authentication of the data signal and ignore the pilot signal. The data signal carries the unitary power binary data stream $d_i$ with symbol period $T_s$ and can be written as
\begin{equation}
p(t)= \sum_i d_i s_p(t-iT_s),
\end{equation}
where
\begin{equation}
\label{eq:spreSym}
s_p(t) \triangleq \sum_{i=0}^{N_c-1} c_i u(t-iT_c)
\end{equation}
is the spreading pulse with chip period $T_c=T_s/N_c$, spreading sequence $c_i=\pm \frac{1}{\sqrt{N_c}} \ i=0, \dots , N_c-1 $  and unitary-energy chip pulse $u(t)$, therefore $p(t)$ has unitary power. In the Galileo system the chip pulse is a sequence of signed rectangular functions with finite support $T_c$ \cite{galileoICT}.

We also propose an improvement of existing systems with the addition of a (public) \textit{authenticated channel} from the ground segment that allows Bob to be sure that messages over this channel are not forged by Eve. 
The information carried by the authenticated channel is available to all users, including Eve. 

Therefore, our model comprises three types of communication channels: the wireless navigation channel from the satellite, the authenticated channel from the ground segment and the attack channel from Eve. 
\subsubsection{Navigation and attack channel}
The navigation channel connects the satellite to users and is the means through which the signal $s_A(t)$ transmitted by Alice propagates. As from Fig. \ref{fig:satelite} we have two navigation channels: one from Alice to Eve, and the other from Alice to Bob. The attack channel connects Eve to Bob and carries the spoofing signal $s_E(t)$.

In an \ac{AWGN} channel, as typically considered in satellite navigation systems, the received signals by Bob and Eve are
\begin{gather}
\label{eq:rBt}
r_B(t)=s_A(t)+s_E(t)+w_B(t), \\
\label{eq:rEt}
r_E(t)=s_A(t)+w_E(t),
\end{gather}
where $w_B(t)$ and $w_E(t)$ are the zero-mean \ac{AWGN} signals with power $\sigb$ and $\sige$, respectively. Moreover, $s_E(t)$ in \eqref{eq:rBt} denotes the attack signal by Eve. Note that \eqref{eq:rEt} may hold at different times, as Eve in general may alternate phases in which she receives the signal from the satellite and transmits the spoofing signal.
In the current standard $s_A(t) = p(t)$.

\subsubsection{Authenticated channel}
We assume that the ground segment can communicate with all the users through an authenticated data channel.  The authenticated channel is assumed to be of large (infinite) bandwidth provided for example through an Internet connection. The authentication is ensured by higher layer authentication protocols \cite{stallings} (such as https). We assume Eve has no control over the information travelling on the authenticated channel and, thus, she can not modify it.
Moreover, as no fine time synchronization is available on the authenticated channel, it is not useful for ranging purposes.

\subsection{Attack Models}
\label{sec:attacks}
Eve's objective is to forge a navigation signal, send it to Bob and let him believe it was transmitted by Alice. in this work we model Eve's behaviour with four attacks \cite{Fea2,FeaGianluca,SCER}

\subsubsection{Forward estimation attack}
Although \ac{NMA} can be used to introduce unpredictability, eventually all data bits will go through channel encoding. Eve can then exploit redundancy to guess the whole codeword by just looking to a fraction of the codeword itself \cite{Fea2}. In this case Eve observes a few symbols $d_i$ and then predicts the rest of the codeword.

\subsubsection{Delay attack}
All bits in the current Galileo navigation message (e.g. ephemeris, navigation data, clock synchronization bits)  are predictable and publicly available to download, therefore the entire codeword is predictable.
In this case Eve knows in advance the signal transmitted by Alice  and can superimpose a powerful time-shifted version $s_E(t)$ of this signal to Alice's signal. Bob will synchronize on the strongest signal and then acquire the timing chosen by Eve. As a consequence Eve will be able to induce the desired (false) ranging on Bob by properly choosing the time shift. 
Assume now that Eve is able to predict the legitimate signal with a delay $\Delta$. Eve can  transmit noise to Bob up to time $\Delta$, and then the properly delayed signal. With this technique $s_E(t)$ can also be chosen in order to remove $s_A(t)$.

\subsubsection{Symbol prediction attack}
In this attack Eve works at the waveform level. The chip pulse $u(t)$  (perfectly predictable \cite{galileoICT}) can be estimated by Eve on a sample by sample basis. Thus, assuming a noiseless reception ($\sige=0$), by reading a small time portion $\Delta$ of $r_E(t)$, Eve can predict the whole symbol.
In the literature this attack is also known as \ac{SCER} attack \cite{SCER} when dealing with ranging signals protected by cryptography (and usually considering $\sige>0$).

\subsubsection{Replay attack}
In the replay attack Eve retransmits to Bob  the received signal instantly, right after reception, with arbitrary power.
Therefore, the replayed signal contains also the non-predictable components $x(t)$ and $w^*(t)$, thus differing from the legitimate signal only by the noise possibly introduced by Eve's front-end.
In this paper we do not analyse the replay attack since we consider the worst case scenario in which $\sige = 0$, therefore the malicious received signal is mathematically undistinguishable from the legitimate one and the replay attack would always succeed (no matter of which authentication procedure is used).

\section{Authentication protocol}
\label{sec:ITA}
The proposed protocol comprises two phases: in the first phase Alice superimposes to the ranging signal $p(t)$ a synchronous authentication signal $x(t)$ carrying a message $V$ and an \ac{AN} signal $w^*(t)$ that prevents the predictive attacks.
Both the AN and the message $V$ are generated by the ground segment and conveyed to Alice through a secure authenticated channel.
In the second phase the AN and the message $V$ are revealed to Bob (and Eve), through the authenticated channel. Bob removes the \ac{AN} from the originally received signal, decodes the authentication message and checks its correspondence with $V$ to confirm the authenticity of the received signal. We now detail the operations carried out in the two phases.

\begin{figure*}
\centering
\includegraphics[width=1.0\textwidth]{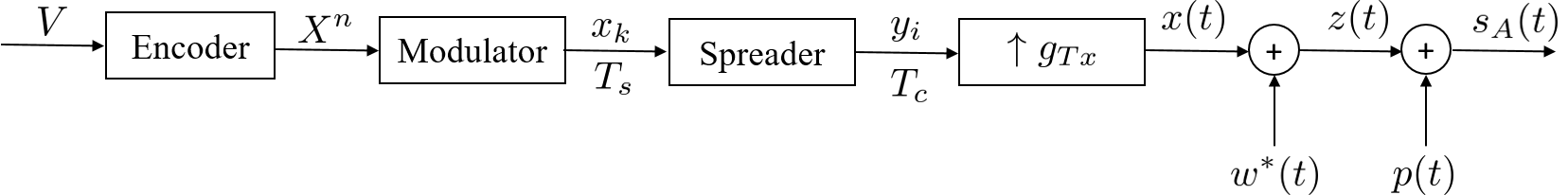}\caption{Transmitter scheme for phase 1.}
\label{fig:txX}
\end{figure*}

\begin{figure*}
\centering
\includegraphics[width=1.0\textwidth]{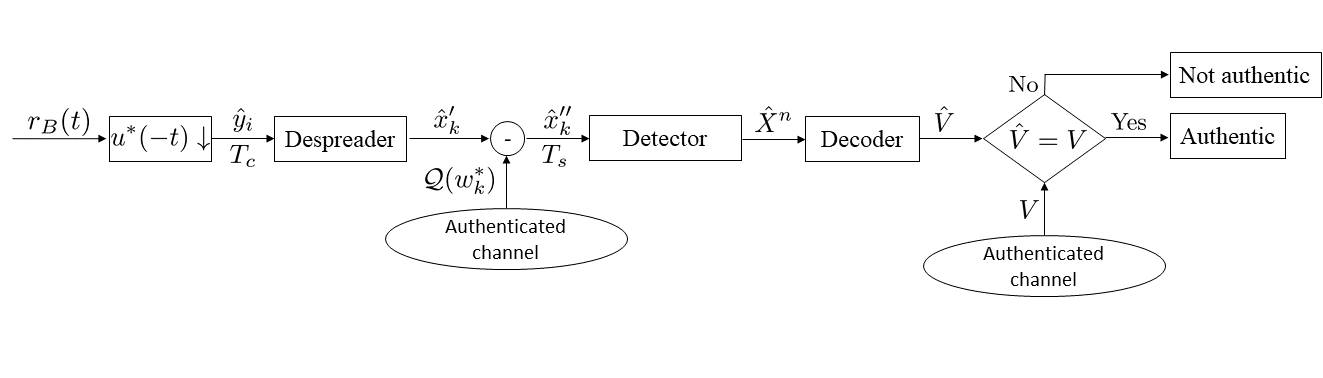}\caption{Bob signal processing for phase 2.}
\label{fig:rxX}
\end{figure*}

\paragraph*{First phase}
In the first phase Alice transmits the authentication signal generated as described in Fig. \ref{fig:txX}. In particular Alice encodes a secret authentication message $V$ in a codeword $X^n$. 
The codeword enters the modulator which outputs constellation symbols $x_k$ with power $\sigx$ at symbol time $T_s$. Then each symbol is spread with the spreading sequence $c_{A,i}= \pm \frac{1}{\sqrt{N_c}}, \ i=1,\dots, N_c$, yielding the $T_c$-sampled signal
\begin{equation}
y_i=x_{\floor{i/N_c}} c_{A,i \text{ mod } N_c}.
\end{equation}
Finally the chip pulse $u(t)$ is used to obtain the continuous time signal
\begin{gather}
x(t)=\sum_i y_i u(t-iT_c).
\end{gather}
In order to guarantee the secrecy of message $V$, we use AN superimposed to $x(t)$, whose power is chosen such that that even if Eve has a noiseless receiver, she cannot decode (and predict) $V$ (see Section \ref{sec:performanceAnalysis} for details). 
The characteristics of $w^*(t)$ will be specified in Section \ref{sec:corrSec}. The signal
\begin{equation}
	z(t)=x(t)+w^*(t)	
\end{equation}
is superimposed to the ranging signal $p(t)$ and the signal transmitted by Alice becomes
\begin{equation}
\label{eq:sA}
s_A(t)=z(t)+p(t).
\end{equation} 
Both authentication and \ac{AN} signals are chosen orthogonal to the ranging signal, \ie the despreading of these signals through the spreading code used for the ranging signal provides a null signal.
This is achieved by using an orthogonal spreading code for the authentication signal and projecting the \ac{AN} on the orthogonal space to the ranging signal, as detailed in Section \ref{sec:corrSec}.
In particular, for the authentication signal  the spreading signal $c_{A,i}$ is chosen to ensure orthogonality with sequence $c_i$ used for $p(t)$.
Therefore, a legacy receiver is not affected by the new superimposed signals. 
 
The signals received by Bob and Eve on the \ac{AWGN} channels are still given by \eqref{eq:rBt} and \eqref{eq:rEt} with the new transmitted signal $s_A(t)$ of \eqref{eq:sA}.
Bob acquires the synchronization on signal $p(t)$, samples and despreads the received signal with sequence $c_{A,i}$ as shown in Fig. \ref{fig:rxX} to obtain the equivalent discrete-time despread signal, which in the absence of attack is  
\begin{equation}
\label{eq:xp}
\hat{x}_k'= z_k+w_{B,k},
\end{equation}
where $z_k=x_k+w^*_k$. The noise samples are still \ac{iid} with zero mean and powers $\sigs$ and $\sigb$ respectively and
\begin{gather}
	w_{k}^*= \int_{kT_s}^{(k+1)T_s} w^*(\tau), s_R(\tau-kT_s) d\tau, \\
w_{B,k}=\int_{kT_s}^{(k+1)T_s} w_B(\tau) s_R(\tau-kT_s) d\tau, \end{gather}
where
\begin{equation}
s_R(t)=\sum_{i=0}^{N_c-1} c_{A,i} u^*(iT_c-t).
\end{equation}
Note that we have omitted in \eqref{eq:xp} the navigation signal component as it is orthogonal to $z(t)$.
Similarly also Eve obtains
\begin{gather}
\label{eq:xEk}
\hat{x}'_{E,k}=z_k+w_{E,k}, \\
w_{E,k}=\int_{kT_s}^{(k+1)T_s} w_E(\tau) s_R(\tau-kT_s) d\tau
\end{gather}
and her \ac{SNR} is 
\begin{equation}
\label{eq:gammaE}
	\Gamma_E=\frac{\sigx}{\sigs + \sige}, 	
\end{equation}
since in the first phase she does not know the \ac{AN}.

\paragraph*{Second phase}
In the second phase 
\begin{enumerate*}[label=\emph{\alph*})]
\item Alice transmits information on $V$ and the \ac{AN} on the authenticated channel and
\item Bob elaborates the signal received in the first phase according to the scheme of Fig. \ref{fig:rxX}.
\end{enumerate*} 
Note that the AN samples $w_k^*$ can be taken from a finite alphabet to simplify their transmission over  the authenticated channel. 
Otherwise, even when $w^*_k$ is a continuous valued random variable, the ground segment quantizes $w^*_k$ into $\mathcal{Q}(w_k^*)$ using $\mathtt{b}$ bits and sends it over the authenticated channel.
The parameter $\mathtt{b}$ must be chosen as a trade-off between performance and cost. 
Here we focus on this latter quantization option.

In the absence of attack and perfect synchronization the quantization error is
\begin{equation}
w_k^{(q)} \triangleq w_k^*-\mathcal{Q}(w^*_k),
\end{equation}
with zero mean and power $\sigq$. In the following we approximate the residual quantization error as Gaussian, as a common practice in the literature. 
Together with $w_k^*$ the ground segment also reveals the original message $V$.

As shown in Fig. \ref{fig:rxX} Bob subtracts from the signal received in phase 1 $r_B(t)$ the quantized \ac{AN} received through the authenticated channel obtaining the signal
\begin{gather}
\label{eq:xsec}
\hat{x}_k''= x'_k - \mathcal{Q}(w_K^*) = x_k + w_{B,k} + w_k^{(q)}.
\end{gather}
Bob detects and decodes the message $\hat{V}$ from the received signal $\hat{x}_k''$. If $\hat{V}=V$ Bob declares that the authentication signal comes from Alice and the ranging signal is also authentic. Otherwise Bob generates an exception and the ranging signal is declared not authentic. 
Since both sample and frame synchronizations are obtained from $p(t)$, we design the signal such that any misalignment between $p(t)$ and $x(t)$ results in an error of the decoded message $V$, thus revealing the attack (see Section \ref{sec:sync}). 
Note that with perfect reconstruction, i.e., without  quantization, $w_k^{(q)}=0$. The resulting \ac{SNR} is 
\begin{equation}
\label{eq:gammaB}
	\Gamma_B=\frac{\sigx}{\sigb+\sigq}.	
\end{equation}

\subsection{Correctness and Security Properties}
\label{sec:corrSec}
The \textit{correctness} of the protocol is the ability to properly authenticate the navigation signal coming from Alice. This happens if $\hat{V}=V$ when Alice is transmitting: we must ensure that $V$ is decodable after phase two, i.e., after the reception of the side signal on the authenticated channel.

The \textit{security} of the proposed protocol is ensured when Eve is not able to decode $V$ in the first phase, and in particular some bits are completely unknown to her (thus having probability 0.5 each of being equal to 0 or 1).
In this case Eve will not be able to generate the authentication signal to deceive Bob.

The two conditions of correctness and security correspond to those of a wiretap transmission scenario \cite{bloch}. Specifically the legitimate received signal is $\hat{x}''_k$ in \eqref{eq:xsec}, over which we require \emph{authenticity}; the malicious received signal is $\hat{x}_{E,k}'$ in \eqref{eq:xEk}, over which we require \emph{secrecy}.

From the correctness and security analysis we obtain the following requirements for the proposed authentication protocol:
\begin{enumerate}
\item \emph{Orthogonality} between $z(t)$ and $p(t)$. For the \ac{AN} we first generate a stationary Gaussian process $w(t)$ with $0\leq t \leq T_s$ and then project on the energy-normalized version of $p(t)$, i.e.
\begin{equation}
w^*(t)=w(t)-\rho \frac{p(t)}{\sqrt{E_{p}}},
\end{equation}
with

\begin{gather}
E_{p}= \int_0^{T_s} p^2(t)dt, \\
\rho=\int_0^{T_s}  w(t) \frac{s_p(t)}{\sqrt{E_{p}}} dt.
\end{gather}
Orthogonality of $x(t)$ and $p(t)$ is instead obtained when spreading sequences $c_i$ and $c_{A,i}$ are orthogonal.
This requirement ensures that the authentication signal does not interfere with the navigation signal, thus not affecting a legacy receiver not implementing the authentication features. For the same reason also $w^*(t)$ must be orthogonal to $p(t)$.
\item \emph{Secrecy} of the message $V$ to Eve. The authentication message must not be known to Eve during the first phase, a condition that can be written as
\begin{equation}
\label{eq:mutInf}
\mathbb{I}(V;r_E(t))=0,
\end{equation}
where $\mathbb{I}(\cdot;\cdot)$ denotes the mutual information function. Note that 
\eqref{eq:mutInf} implies also $\mathbb{I}(V;r_B(t))=0$ when Eve has a better channel than Bob ($\sige < \sigb$). In this way we guarantee that in the first phase Eve is not able to correctly decode $x(t)$ \cite{bloch} as she operates above the channel capacity, thus preventing prediction attacks.
\item \emph{Authenticity} in the second phase. We must ensure that Bob is able to decode $\hat{V}$ in the second phase and match it with $V$, received through the authenticated channel, \ie
\begin{equation}
P_e^B \triangleq \mathbb{P}[\hat{V}\neq V
| \text{no attack}]=0,
\end{equation}
where $\mathbb{P}[\cdot]$ denotes the probability.
\item \emph{Synchronization}. Since $x(t)$ and $p(t)$ are orthogonal, Eve can always distinguish between the two messages and operate a predictive attack on $p(t)$. She can delay or anticipate $p(t)$ without interfering with the authentication procedure. We must then require $x(t)$ to be synchronized to $p(t)$ so that a delay in the latter would reveal the attack.
\end{enumerate}
Secrecy and synchronization requirements deal with the \emph{security} metric of the protocol since they both aim at maintaining $V$ secret from Eve. Reliability and authenticity instead deal with correctness since they are the necessary conditions to let Bob properly decode $V$ and authenticate $p(t)$.

\paragraph*{Remark}
An alternative formulation of the authentication protocol does not require the feedback of the message $V$ through the authenticated channel.
In this case Bob decodes the received codeword and decides on the authenticity of $\hat{V}$ based on a threshold over the soft information output of the decoder. 
Note in fact that for a forged authentication signal, the cancellation of the \ac{AN} (still provided through the authenticated channel) would actually add significant noise (as Eve cannot use the correct \ac{AN}), thus making the decoding hard. Therefore, if the likelihood (provided by the decoding algorithm) of $\hat{V}$ is below the threshold, Bob declares $\hat{V}$ as not authentic, since it may have been the result of a guessing attack by Eve. 
This approach, however, requires  the optimum threshold level to ensure given false alarm and missed detection probability values, which will depend also on the length of the codewords.

Another possibility is to apply hypothesis testing to the signal received by Bob discriminating between the legitimate received signal and the received signal under spoofing attack \cite{eusipco}.
Also this approach requires to find the optimum threshold level given a false alarm and miss detection probability values.

With the authentication protocol presented in this paper, instead, there is no need for a threshold, since the authenticity check is performed simply by checking if the two messages $V$ and $\hat{V}$ are equal.

\section{Protocol performance analysis}
\label{sec:performanceAnalysis}
We now analyse both the correctness and the security of the proposed algorithm against various kinds of attacks. In particular, we consider \ac{FEA} under the hypothesis of both  infinite-length and finite-length codewords. We analyse also a simple attack in which Eve only replaces the navigation signal with a spoofed delayed one, breaking the synchronization between the authentication and the navigation signals.

\subsection{Forward Estimation Attack - Infinite-length Codewords}
\label{sec:FEAinf}
We consider here the forward estimation attack with ideal signalling, \ie when codewords have infinite lengths and a real Gaussian modulation is used for $x_k$. Let $R_x$ be the code rate of $x_k$. From results on wiretap-coding we have that secrecy condition \eqref{eq:mutInf} is satisfied as long as \cite[Chaper~5]{bloch}
\begin{equation}
\label{eq:sec}
	R_x \geq C_E= \frac{1}{2} \log_2 \left( 1+\Gamma_E \right).	
\end{equation}
Similarly, the authenticity is ensured as long as Bob is able to decode $V$ in the second phase, \ie 
\begin{equation}
\label{eq:rel}
	R_x \leq C_B= \frac{1}{2}	\log_2 \left(1+\Gamma_B\right),
\end{equation}
where $\Gamma_B$ is given by \eqref{eq:gammaB}.
Therefore, assuming as worst case that Eve has a noiseless receiver ($\sige = 0$), from \eqref{eq:gammaE} and \eqref{eq:sec} the noise power $\sigs$ must satisfy
\begin{equation}
	\sigs \geq \frac{\sigx}{2^{2R_x}-1} .	
\end{equation} 
$C_E$ and $C_B$ are the channel capacities of Eve and Bob respectively.
Note that the additional information in the second phase ($\mathcal{Q}(w_k^*)$ and $V$) must be transmitted over the authenticated channel to prevent Eve from altering its content and matching her counterfeit signal.

Eve can still attempt to predict the codeword, by guessing the secret bits that are unknown to her. By the wiretap coding theory, there exist suitable wiretap codes for Alice such that the part of the authentication message that remains secret to Eve has a secrecy rate
\begin{equation}
	R_A=R_x-C_E,	
\end{equation}
which is maximized when $R_x=C_B$ and we obtain the secrecy capacity \cite{bloch}
\begin{equation}
	C_A\triangleq C_B-C_E.	
\end{equation}
Note that in our context the secrecy of message $V$ is only instrumental to the authentication of the navigation message. Therefore, with a small abuse of notation, we will denote as \textit{authentication capacity} the secrecy capacity $C_A$, as the secret bits are those that prevent Eve from obtaining a successful attack.

The probability that Eve predicts the correct message $V$ is
\begin{equation}
	P_{\text{succ}}	=2^{-R_A\bar{n}},
\end{equation}
where $ \bar{n} \to \infty$ is the codeword length (in symbols). 
When finite-size constellations are considered for $x_k$, conditions \eqref{eq:sec} and \eqref{eq:rel} still hold for correctness and secrecy. 
However, $C_E$ becomes the achievable rate of a finite-size constellation system on an \ac{AWGN} channel with \ac{SNR} $\Gamma_E$, \ie

\begin{equation}
C_E=\mathbb{H}(y)-\log_2 \left(\frac{\pi e}{\Gamma_E} \right)
\end{equation}
with $\mathbb{H}(y)$ being the entropy of the received signal with \ac{PDF} $f_y(a)$
\begin{gather}
\label{eq:yenthropy}
\mathbb{H}(y)=\int_{\mathbb{C}} f_y(a)\log_2 \frac{1}{f_y(a)} da, \\
f_y(a)=\frac{1}{M}\sum_{s \in \mathcal{S}} \frac{\Gamma_E}{\pi } e^{-|s-a|^2\Gamma_E},
\end{gather}
and $\mathcal{S}$ is the set of the $M$ complex constellation points. 
In order to compute the capacities we must resort to the numerical integration of \eqref{eq:yenthropy}. A similar expression holds for $C_B$ where $\Gamma_E$ is replaced by $\Gamma_B$. 


\subsection{Forward Estimation Attack - Finite-length Codewords}
\label{sec:FEAfin}
The previous section provided an analysis for the scenario of infinitely long codewords, as an asymptotic performance limit. Here we consider a more realistic scenario of finite-length codewords. We still first assume Gaussian signalling. Due to the finite-length regime, \eqref{eq:sec} and \eqref{eq:rel} do not hold anymore. For correctness we must assess the (non-zero) probability that Bob does not decode $V$, while for secrecy we must assess the (non-zero) probability that Eve correctly predicts $V$ before the entire codeword  has been transmitted. In order to compute these probabilities we resort to literature results on finite block-length regime \cite{erseghe1,erseghe2}. Let us assume the codebook comprises $\gamma$ codewords, that are transmitted with equal probability.
In particular we lowerbound the codeword error probability $P_e\left(\Gamma,\frac{\log_2 \gamma}{\bar{n}},\bar{n}\right) $ on \ac{AWGN} channel with \ac{SNR} $\Gamma$ as
\begin{equation}
\label{eq:bound}
	P_e\left(\Gamma,\frac{\log_2 \gamma}{\bar{n}},\bar{n}\right) \geq q\left(\Gamma,\frac{\log_2 \gamma}{\bar{n}},\bar{n}\right),
\end{equation}
where
\begin{gather}	
	q(\Gamma,R,\bar{n}) \triangleq Q \left( \sqrt{\frac{\bar{n}}{G}} \left( \frac{F-R}{\log_2 e} + \frac{\ln(\bar{n})}{2\bar{n}}  \right)   \right),	\\
	F=\frac{1}{2} \log_2 \left( 1+\Gamma \right), \\
	G=\frac{\Gamma (2+\Gamma)}{2(1+\Gamma)^2},
\end{gather}
and $Q(\cdot)$ is the complementary \ac{CDF} of a continuous normal variable.
For a given length $\bar{n}$ a design criterion in this scenario is to set a desired correctness outage probability $\Pi_0$, \ie choose the number of codewords $\gamma$ such that
\begin{equation}
	P_e\left(\Gamma_B,\frac{\log_2 \gamma}{\bar{n}},\bar{n}\right)	< \Pi_0.
\end{equation}
Then, considering a codeword predictive attack performed by Eve at symbol $n<\bar{n}$, the probability of successful attack is upper-bounded as
\begin{equation}
\label{eq:psuccn}
\begin{split}
P_{\text{succ}}(n) &\leq \max \left\lbrace 1- P_e\left( \Gamma_E,\frac{\log_2 \gamma}{n},n \right),\frac{1}{\gamma} \right\rbrace \\ &\leq \max \left\lbrace 1- q\left( \Gamma_E,\frac{\log_2 \gamma}{n},n \right),\frac{1}{\gamma} \right\rbrace,	
\end{split}	
\end{equation} 
where the second inequality comes from two facts: 
\begin{enumerate*}[label=\emph{\alph*})]
\item $q(\Gamma,\frac{\log_2 \gamma}{n},n)$ is a lower bound on the codeword error probability and
\item equation \eqref{eq:bound} is based on the fact that the code is optimized for length $\bar{n}$, while Eve attempts the decoding after receiving $n$ samples, thus we have a further source of error by this mismatch.
\end{enumerate*}
The maximum comes from the fact that the success probability cannot be lower than $1/\gamma$, which corresponds to the complete random choice of the attack codeword.

We now consider the impact of finite-size constellations, and in particular we consider a binary modulation. For this case  \eqref{eq:bound} still holds with \cite{erseghe1,erseghe2}
\begin{gather}
	F=1+\frac{H^{(1)}}{\ln(2)}, \\
	G = H^{(2)} - (H^{(1)})^2,
	\end{gather}
	where
	\begin{gather}
	H^{(\ell)}=\frac{1}{\sqrt{2\pi \Gamma}}
		\int_{-\infty}^{\infty} e^{-\frac{1}{2\Gamma}(b-\Gamma)^2}
		(-h(b))^\ell db, \\
	h(b)=\ln (1+e^{-2b}).
\end{gather}

Also in this case, the inability of Eve to predict the \ac{AN} further lowers the success of the attack, as the spoofed \ac{AN} will not be completely removed by Bob before decoding of the authentication message.
\subsection{Delay Attack}
\label{sec:sync}
We now consider the delay attack, in which Eve does not attempt to reproduce the authentication signal, but only transmits a delayed navigation signal. Assuming that Bob acquires the synchronization on the spoofed signal, \ie the attack is successful, we aim at assessing the probability that Bob also demodulates $V$ from the asynchronous authentication signal, thus failing to reveal the attack. Let $-T_s<\epsilon<T_s$ be the offset between the navigation and the authentication signals, \ie in phase one
\begin{equation}
	r_B(t)=p(t)+z(t-\epsilon)+w_B(t).	
\end{equation}
First consider the case $0\leq\epsilon <T_s$. After despreading and \ac{AN} removal $\hat{x}''_k$ in \eqref{eq:xsec} is affected by the previously transmitted symbol $x_{k-1}$, \ie
\begin{equation}
\label{eq:xInterf}
		\hat{x}''_k= \alpha x_k + \beta x_{k-1} + w_{B,k} + w_{k,\epsilon}^{(q)},
\end{equation}
where the interference coefficients are
\begin{gather}
\alpha = \int_{\epsilon}^{T_s} s_T(\tau-\epsilon) s_R(\tau) d\tau, \\
\beta= \int_ {0}^{\epsilon} s_T(\tau+T_s - \epsilon) s_R(\tau)d\tau, \\
s_T(t)=\sum_{i=0}^{N_c-1} c_{A,i} g_{Tx}(t-iT_c). 
\end{gather}
In \eqref{eq:xInterf}, besides the inter-symbol interference there is also the residual quantization error $w_{k,\epsilon}^{(q)}$ that now depends on the delay $\epsilon$. In particular we have

\begin{equation}
	w_{k,\epsilon}^*= \int_{kT_s}^{(k+1)T_s} w^*(\tau-\epsilon) s_R(\tau-kT_s)	d\tau,
\end{equation}
and thus
\begin{equation}
\label{eq:wkstar}
	w_{k,\epsilon}^{(q)}=w_{k,\epsilon}^* - \mathcal{Q}(w_k^*).	
\end{equation}
The power of $w_{k,\epsilon}^{(q)}$ is
\begin{equation}
\sigqe(\epsilon) = \E{|w_{k,\epsilon}^{(q)}|^2},		
\end{equation}
where $\E{\cdot}$ is the expectation operator.
Considering perfect quantization, \ie $w_k^*=\mathcal{Q}(w_k^*)$, $w_{k,\epsilon}^*$ and $w_k^*$ are two correlated Gaussian random variables. Note that
\begin{equation}
	\sigqe=\E{(w_{k,\epsilon}^{(q)})^2} + \E{(w_k^*)^2} - 2 \E{w_{k,\epsilon}^{(q)} w_k^*}. 	
\end{equation}
 Now we have 
\begin{equation}
\label{eq:crossCorr}
	 \begin{split}
	 	 &\E{w_{k,\epsilon}^* w_k^*} = \\ 
	 	 &=\E{ \int_0^{T_s} \int_0^{T_s}  w^*(\tau)
	 	s_T(\tau) w^*(\tau'-\epsilon) s_R(\tau')     d\tau' d\tau } \\
	 	& = \int_0^{T_s} \int_0^{T_s} \E{w^*(\tau)w^*(\tau'-\epsilon)} s_T(\tau) s_R(\tau')  d\tau' d\tau,
	 \end{split}
\end{equation}
where, the second line comes from \eqref{eq:wkstar}, the third line comes from the linearity of the expectation and we considered $k=0$ in the integral limits for the noise stationarity.
Since $w^*(t)$ is a white Gaussian process, by definition the inner expected value becomes
\begin{equation}
\label{eq:delta}
	\E{w^*(\tau)w^*(\tau'-\epsilon)}= \delta (\tau-\tau'+\epsilon) \sigs,	
\end{equation} 
where $\delta(\cdot)$ is the continuous time impulsive function. Due to the integral properties of $\delta(\cdot)$ \eqref{eq:crossCorr} becomes 
\begin{equation}
		\E{w_k^\epsilon w_k^*}=\sigs \int_0^{T_s-\epsilon} 
		 s_T(\tau) s_R(\tau+\epsilon) d\tau =\sigs\nu_\epsilon, 
\end{equation}
where the result of the integral $\nu_\epsilon$ only depends on $\epsilon$ and the transmitter and receiver pulses. Note that if $\epsilon=0$, then $w_k^*=w_{k,\epsilon}^*$ and $w_{k,\epsilon}^q=0$. Moreover, for a high $\epsilon$ the correlation between $w_k^*$ and $w_{k,\epsilon}^*$ decreases; if $\epsilon$ exceeds $T_s$ the two variables become uncorrelated ($\nu_\epsilon=0$), since they insist on disjoint intervals of $w^*(t)$.
Under these conditions $\sigqe=2\sigs (1-\nu_{\epsilon})$ and the \ac{SNR} becomes

\begin{equation}
	\Gamma'_B=\frac{\alpha^2 \sigx}{ \beta^2 \sigx+ \sigb+\sigqe}.	
\end{equation}
Note that if there is no delay, i.e. $\epsilon=0$, we have $\alpha=1$, $\beta=0$, $\sigqe=0$ and hence $\Gamma_B'=\Gamma_B$. If, on the other hand, $\epsilon>0$, then $\alpha<1$ and $\beta>0$. This, together with $w_{k,\epsilon}^{(q)}$, decreases Bob's \ac{SNR} and mines his capability to decode $\hat{V}$, resulting in the attack being uncovered.

\subsection{Symbol Prediction Attack}
With the symbol prediction attack Eve aims at detecting the symbol transmitted by Alice in order to send a delayed version of it. Due to the presence of the \ac{AN} the prediction of the authentication message symbols is more difficult for Eve. In particular, since even the detection of the whole codeword will be affected by a codeword error rate bounded away from zero when operating above the authentication capacity, (by the converse of the wiretap channel coding Theorem)
detection at symbol level will also be affected by errors, or otherwise the concatenation of correctly detected symbols would provide the correct codeword. In the case of a binary constellation the success probability upon a symbol prediction made at time $0 < t\leq T_s$ is \cite{FeaGianluca}
\begin{equation}
\label{eq:scerSucc}
	P_{succ}=1-Q\left( \sqrt{2\Gamma_E  \frac{t}{T_s}} \right).
\end{equation}
Also in this case the \ac{AN} spoofed by Eve in the predicted part of the symbol will be added to $w_k^{(q)}$ at Bob and authentication message decoding will fail.

\section{Numerical Results}
\label{sec:results}

We consider the transmission scenario of Fig. \ref{fig:satelite} with a single satellite, where all satellite links are modelled as \ac{AWGN} channels. The authenticated channel has been assumed error-free and with a large band (we will also consider the effects of noise quantization). As for the Galileo signal we assume $N_c=4,092$ and $T_c = 10^{-6} / 1.023$ \cite{galileoICT}. We focus on a unitary-power authentication signal, \ie $\sigx=1$, while different values for the \ac{AN} power will be considered. For Bob's noise power we set $\sigb=0,-5,-10$ dB, values typically encountered in GNSS receivers \cite{snrTypical}. For Eve, we assume $\sige=0$ as a worst case for the authentication problem, corresponding to a noiseless receiver.
\begin{figure*}
\centering
\subfloat[Transmission chip pulse of the Galileo standard.]{\label{fig:standard} \includegraphics[width=0.5\textwidth]{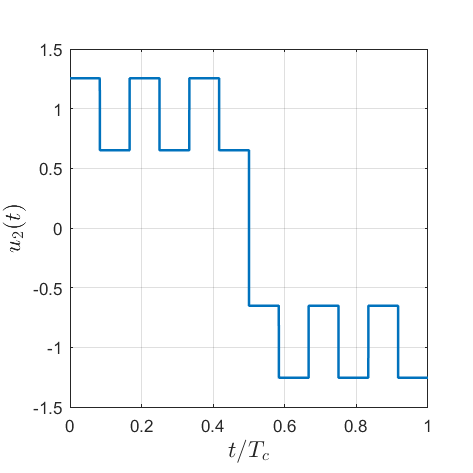}}
\subfloat[Proposed transmission chip pulse $u_2(t)$.]{\label{fig:CsDegGaussHornsC} \includegraphics[width=0.5\textwidth]{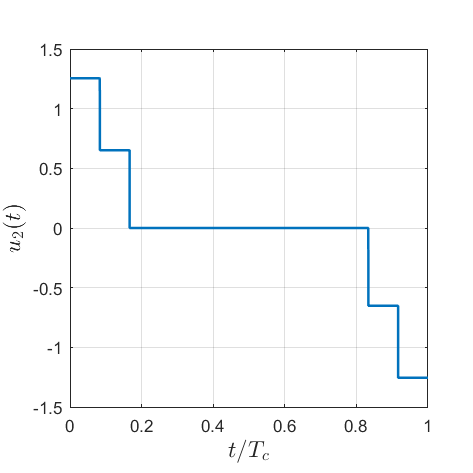}}
\caption{Two considered chip pulses.}
\label{fig:twoPulses}
\end{figure*}
As transmission chip $u(t)$ we consider two options, shown in Fig. \ref{fig:twoPulses}. In particular $u_1(t)$ is the chip pulse used in the Galileo system \cite{galileoICT}, while $u_2(t)$ is a chip pulse characterized by a smaller support designed in order to make the authentication signal more fragile to synchronization errors, as discussed in Section \ref{sec:sync}.
Results in this section are based on the analysis presented in the paper with the \ac{AWGN} channel described in Section \ref{sec:SM}. For a practical implementation of the system further improvements should be considered, which are left for future work.

\begin{figure}
\centering
\includegraphics[width=0.5\columnwidth]{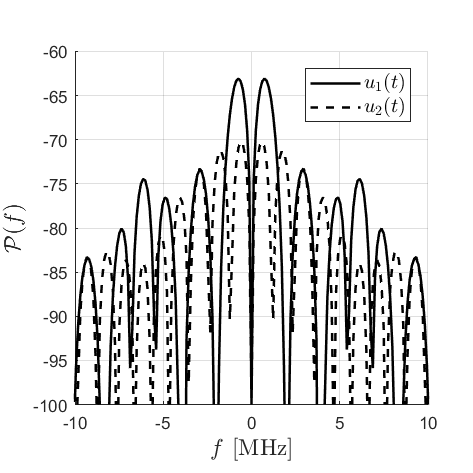}
\caption{PSD of $x(t)$ modulated by the two different chip pulses of Fig. \ref{fig:twoPulses}.}
\label{fig:PSDs}
\end{figure} 
Fig. \ref{fig:PSDs} shows the \ac{PSD} $\mathcal{P}(f)$ of the chip pulses. We note that the new pulse has a similar \ac{PSD} to the standard one. Still the design of the pulses involves many other issues and we do not aim to propose new solutions here; instead we only consider $u_2(t)$ as an example of other possible pulses that are promising  for authentication purposes.

\subsection{Forward Estimation Attack - Infinite Codeword Length}
\begin{figure}
\centering
\includegraphics[width=0.5\columnwidth]{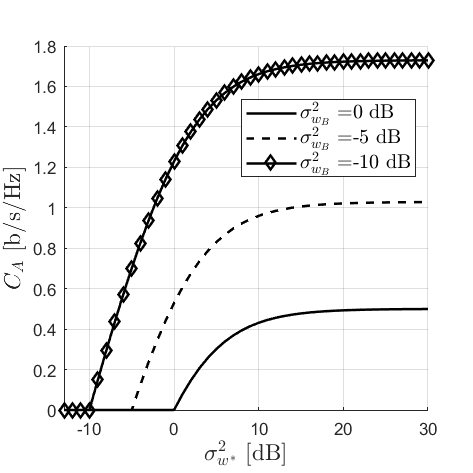}\caption{Authentication capacity versus $\sigma^2_{w^*}$ for three different values of $\sigb$.}\label{fig:CsVsWstar}
\end{figure}
\begin{figure}
\centering
\includegraphics[width=0.5\columnwidth]{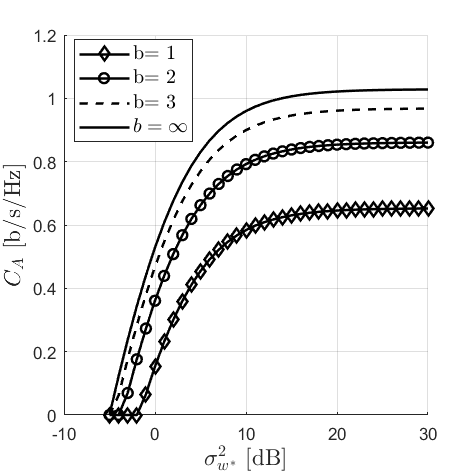}
\caption{Authentication capacity versus $\sigs$ for some values of quantization bits $\mathtt{b}$ and $\sigb=-5$ dB.}
\label{fig:quantization}
\end{figure}

We first consider FEA with infinite codeword length, as analysed in Section \ref{sec:FEAinf}. Moreover we consider infinite-rate authenticated channel, thus $\sigqe=0$. Fig. \ref{fig:CsVsWstar}  shows the secrecy capacity as a function of the \ac{AN} power $\sigs$ for different values of Bob's noise power. We observe that the capacity is zero for $\sigs$ below the threshold, $\sigs \leq \sigb$, and then increases with $\sigs$. Moreover, as $\sigs$ goes to infinity, the secrecy capacity saturates to the Alice-Bob channel capacity (as Bob's noise is limiting the capacity anyway). For example at $\sigs=0$ dB, \ie with \ac{AN} having the same power of the authentication signal (and of the navigation signal) we have $C_A=0.52$ b/s/Hz  for $\sigb=5$ dB. Note that this choice would require the reduction of the navigation signal power by 4.7 dB for the same total satellite transmit power. We also considered the \ac{AN} quantization in the authenticated channel. In particular, we consider a uniform quantizer optimized in order to minimize the mean square quantization error \cite{dataCompression}. Fig. \ref{fig:quantization} shows the authentication capacity as a function of $\sigs$  and as a function of the number of quantization bits per sample ($\mathtt{b}$)  for $\sigb=-5$ dB.
We also include the performance for the case of no quantization error ($\mathtt{b}=\infty$). We observe that already with $\mathtt{b}=3$ the authentication capacity loss is below 0.3 b/s/Hz.

\begin{figure}
\centering
\includegraphics[width=0.5\columnwidth]{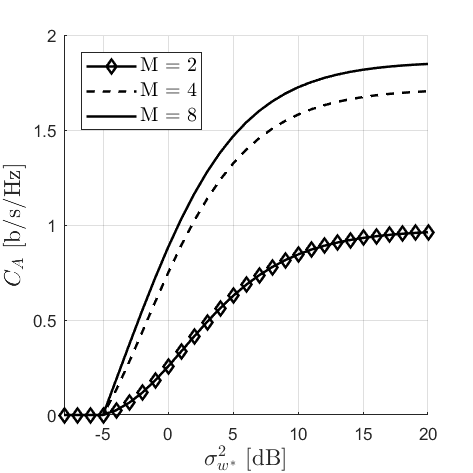}
\caption{Authentication capacity of the $M$-PSK constellation versus $\sigs$ for $\sigb=-5$ dB.}
\label{fig:mpsk}
\end{figure}

For the case of finite constellations, we focus on the constant-envelope \ac{PSK} with $M$ symbols. Fig. \ref{fig:mpsk} shows the authentication capacity vs the \ac{AN} power for various values of $M$ and $\sigb=-5$ dB. In this case the curves saturate at the constellation-constrained capacity of the Alice-Bob channel and also the \ac{AN} power value yielding zero authentication capacity is now depending on $M$.

\subsection{Forward Estimation Attack - Finite Codeword Lengths}
\begin{figure}
\centering
\includegraphics[width=0.5\columnwidth]{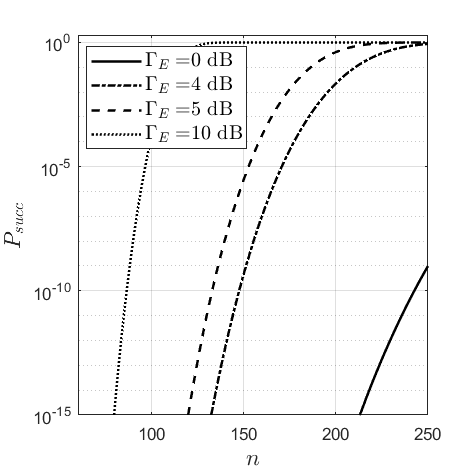}
\caption{\ac{FEA} success probability for $\sigb=-5$ dB, different values of $\Gamma_E$ and Gaussian modulation.}
\label{fig:feaGauss}
\end{figure}
We now consider the case of finite-length codewords as described in Section \ref{sec:FEAfin}. We consider here $\sigb=-5$ dB and no quantization error. For the correctness, by imposing a decoding outage probability of $\Pi_0=10^{-3}$ we obtain from the bound \eqref{eq:bound} $\gamma=1.1 \cdot 10^{64} $. We also chose $\bar{n}=250$, which corresponds to the codeword length of the Galileo \ac{FEC}.

\begin{figure}
\centering
\includegraphics[width=0.5\columnwidth]{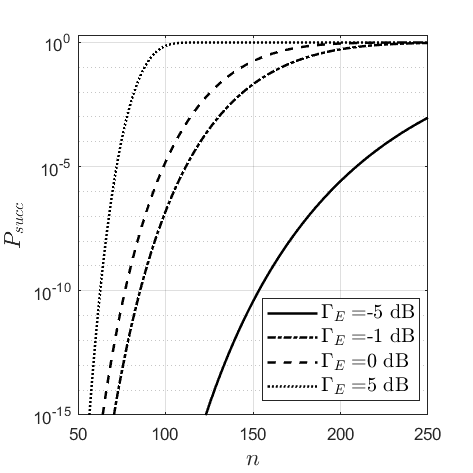}
\caption{FEA success probability versus $n$ for $\sigb=0$ dB, different values of $\Gamma_E$ and BPSK modulation.}
\label{fig:feaBin}
\end{figure}

Fig. \ref{fig:feaGauss} shows the attack success probability $P_{\text{succ}}$ vs the attack delay $n$ and different values of Eve's SNR using \eqref{eq:psuccn}. In this figure we can distinguish between the performance in case we use the proposed authentication protocol or not. If we introduce the \ac{AN} and assume, as a worst case scenario, $\sige=0$, then $\Gamma_E=\frac{1}{\sigs}$:
the solid and dash-dotted lines of Fig. \ref{fig:feaGauss} describe the case in which with $\ac{AN}$ we can force $\Gamma_E<\Gamma_B$.
On the other hand if we do not use the authentication protocol, $\sigs=0$ and $\sige>0$, yielding $\Gamma_E=\frac{1}{\sige}$: the dashed and dotted lines shows that Eve can predict (with probability 1) the codeword well before the last symbols when Gaussian signaling is used for the authentication message.
This represents the asymptotic performance that can be obtained optimizing the modulation.
Note that we obtain a much lower $P_{succ}$ probability with respect to that reported in Fig. \ref{fig:feaBin} for BPSK constellations and the same value of $\Gamma_E$ and $n$, even for a higher $\sigb$ ($\sigb = 0$ dB in Fig. \ref{fig:feaBin} and $\sigb = -5$ dB in Fig. \ref{fig:feaGauss})  

\begin{figure}
\centering
\includegraphics[width=0.5\columnwidth]{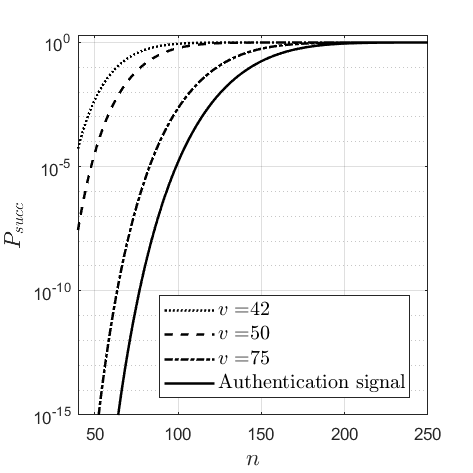}
\caption{FEA success probability comparison between the proposed authentication scheme and \ac{NMA} for different values of $v$, $\Gamma_B = 0$ dB and $\Gamma_E = 0$ dB.}
\label{fig:confrontoFea}
\end{figure}

We then compare the performance of our scheme (solid line of Fig. \ref{fig:confrontoFea}) with an \ac{NMA} approach. 
Let $v$ be number of unpredictable bits in the \ac{NMA} codeword.
We can then use the same formulation \eqref{eq:psuccn} with
$\gamma = 2^v$ and BPSK modulation.
Since with \ac{NMA} there is no \ac{AN} we focus on the case $\Gamma_B = \Gamma_E = 0$ dB.
Note that when considering NMA $\Gamma_E = 1/\sige$, while with our authentication protocol $\Gamma_E =1  / \sigs$.
Fig. \ref{fig:confrontoFea} shows the success probability as a function of $n$ and $v$. Note that $v=42$ (dotted line) is the value considered in \cite{FeaGianluca}. 
Our scheme outperforms \ac{NMA} under FEA.
Moreover, \ac{NMA} performance improves as $v$ increases, at the cost of adding more unpredictable bits, thus increasing the overhead.

\begin{figure*}
\centering
\subfloat[Authentication capacity versus delay $\epsilon$ using pulse $u_1(t)$.
]{\label{fig:standardEpsilon} \includegraphics[width=0.5\textwidth]{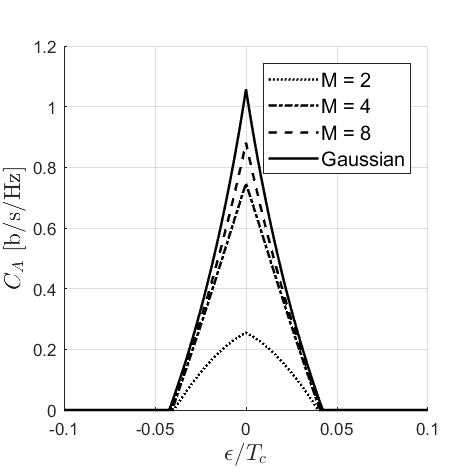}}
\subfloat[Authentication capacity versus delay $\epsilon$ using pulse $u_2(t)$.]{\label{fig:2HornEpsilon} \includegraphics[width=0.5\textwidth]{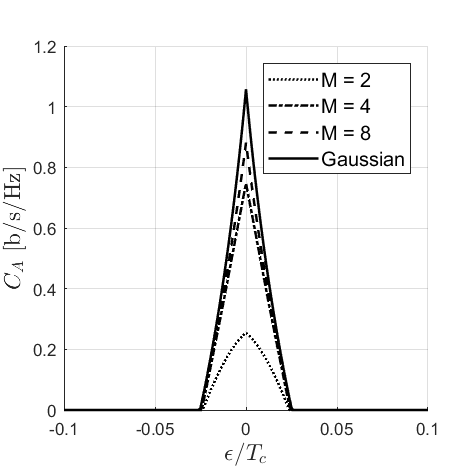}}
\caption{Degradation of authentication capacity versus the delay $\epsilon$ using the transmission chip pulses of Fig. \ref{fig:twoPulses},  $\sigb=-5$ dB and $\sigs=0$ dB.}
\label{fig:epsilon}
\end{figure*}

\subsection{Delay Attack}
For the delay attack we consider the analysis of Section \ref{sec:sync}. 
In particular we consider as \ac{AN} power $\sigs=0$ dB and Bob's noise power $\sigb=-5$ dB. Coding is performed with codewords of infinite length and both Gaussian and $M$-PSK constellations are considered.
Fig. \ref{fig:epsilon} shows the secrecy capacity vs the attack delay $\epsilon$ for various sizes $M$ of the \ac{PSK} constellation , and for the two chip pulses $u_1(t)$ and $u_2(t)$. We observe that for the chip $u_1(t)$ of the Galileo system, the capacity drops to zero for $\epsilon=0.04 \ T_c$, while the pulse $u_2(t)$, having a more compact support, exhibits a zero secrecy capacity already for $\epsilon=0.025 \ T_c$ thus providing a better protection against the symbol prediction attack.

Note that by setting the secrecy coding rate $R_s$ below $C_s(\epsilon^*)=0$ we have that an attack with delay $\epsilon>\epsilon^*$ is detected as, from the converse theorem on capacity, the codeword error probability of Bob tends to 1 as $\bar{n}$ tends to infinity. A suitable choice of the secrecy coding rate $R_s$ takes into account the synchronization error statistics of Bob's receiver in the absence of an attack (done for example to the receiver's noise).

\subsection{Symbol Prediction Attack}
For the symbol prediction attack, the success probability is given by \eqref{eq:scerSucc}. For a non-authenticated signal the SNR is $ 1/\sige$ while for an authenticated signal the SNR of the authentication message is $\Gamma_E$ of \eqref{eq:gammaE}. With $\sige=-5$ dB and $t/T_s=0.3$ (\ie by listening a fraction of the transmission symbol) we have $P_{succ}=0.916$ and $P_{succ}=0.7502$ in the two cases of $\sigs=0$ (no authentication) and $\sigs=0$ dB (with authentication). Again, we note that \ac{AN} significantly lowers the possibility of predicting the authentication message by Eve.

\section{Conclusions}
\label{sec:conclusions}
In this work we proposed a novel authentication protocol and we showed that the proposed solution effectively authenticates a single-satellite navigation message. We analysed the protocol performance under various transmission constraints, such as finite-length codewords, finite-size constellations and quantization. 
We conclude that the proposed strategy is effective in providing authentication of the Galileo signal, totally preventing prediction attacks for Gaussian constellations and significantly lowering the success of attacks for finite-length constellations. Moreover, the unpredictability of the AN further increases the security level of the proposed protocols.

\clearpage

\end{document}